\documentclass{ws-procs9x6}

\usepackage{graphicx}

\begin{document}

\title{STAR Spin related future detector upgrades}

\author{F. Simon (for the STAR Collaboration)}

\address{Massachusetts Institute of Technology \\
77 Massachusetts Avenue \\ 
Cambridge, MA 02139, USA\\ 
E-mail: fsimon@mit.edu}
\maketitle

\abstracts{The STAR experiment at the Relativistic Heavy Ion Collider (RHIC) has a rich spin physics program aimed at exploring the spin structure of the proton with polarized proton beams. In addition to the currently accessible channels, heavy flavor, charged vector boson production and forward mesons are integral parts of the long-term program. Such measurements require upgrades of the STAR tracking system and calorimetry. We are presenting an overview of the planned upgrades and the physics objectives driving them.}

\section{Introduction and Current Capabilities}
The Relativistic Heavy Ion Collider RHIC is the first polarized high-energy proton-proton collider, providing polarized p+p collisions at energies up to $\sqrt{s}$ = 500 GeV. The focus of the polarized p+p program is to study the spin structure of the proton\cite{SpinRev}. This program will provide precise measurements of the polarization of the gluons and of $\bar{u}, \bar{d}, u,$ and $d$ quarks. Transverse spin effects are being explored with transversely polarized beams. 

STAR\cite{STAR} is one of the two large detector systems at this accelerator. Its main tracking detector is a large-volume TPC covering the pseudorapidity range $\vert \eta \vert < 1.2$, with additional vertex resolution for the reconstruction of strange particles provided by the silicon vertex tracker (SVT, $\vert \eta \vert < 1$), and forward tracking by the forward TPCs (FTPCs, $ 2.5 < \vert \eta \vert < 4.0$). The barrel (BEMC) and endcap (EEMC) electromagnetic calorimeter cover $ -1 < \eta <2$, additional small acceptance coverage at high rapidity is provided by the forward pion detector (FPD) in $3.1 < \eta < 4.2$.

\begin{figure}
\centering
\includegraphics[width = 0.99\textwidth]{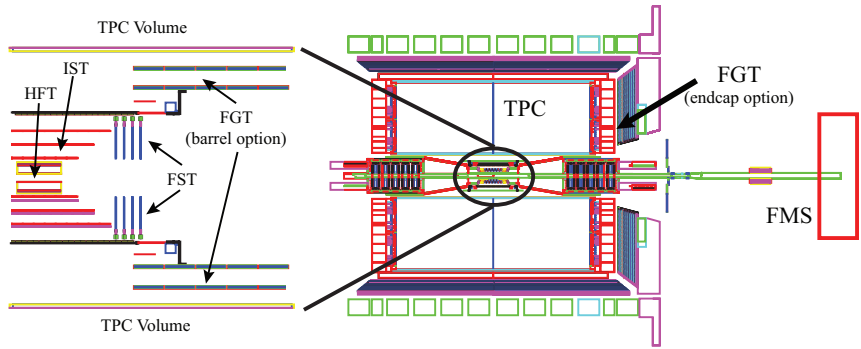}
\caption{Side view of the STAR detector with the planned upgrades. The inner tracking region is shown enlarged. Two possible configurations for the FGT are shown, see text for details.}
\label{fig:Upgrades}
\end{figure}

Due to its large acceptance tracking and electromagnetic calorimetry, STAR is uniquely capable of full jet reconstruction at RHIC and very well suited for measurements of spin asymmetries in inclusive jet, hadron and photon production as well as in di-jet and $\gamma$-jet events. Transverse spin asymmetries at large $x_F$ have been observed in the FPD\cite{Pi0Trans}. Other measurements, such as large acceptance forward measurements with transversely polarized beams, studies of the gluon polarization via the double longitudinal spin asymmetry in heavy flavor production and the flavor separation of quark and anti-quark polarizations via parity violating longitudinal single spin asymmetries require upgrades of the current detector setup. Figure \ref{fig:Upgrades} shows an overview of the planned upgrades within the STAR detector. Details of these upgrades and the physics objectives that drive these upgrades are discussed in the following. 

\section{Heavy Flavor Production}

In hadronic collisions, the production of heavy quark pairs is dominated by gluon-gluon fusion, $gg \rightarrow c\bar{c},\, b\bar{b}$. The double longitudinal spin asymmetry of heavy quark production thus provides direct access to the gluon polarization in the proton and is largely independent of the quark helicities\cite{SpinRev}. In heavy-ion collisions at RHIC, heavy quarks are crucial to investigate the degree of thermalization in early stages of the collisions and can be used to study the properties of the created medium. For these studies it is crucial to identify mesons containing heavy quarks on an event-by-event basis. Since $c\tau\,\sim 120\ \mu$m for $D^0$ and $c\tau\,\sim 460\ \mu$m for $B^0$, very good vertex resolution is needed to identify these particles via displaced vertices. In order to achieve this, an upgrade of the STAR vertex detectors is planned, consisting of two devices, both covering $\vert \eta \vert < 1.0$. The Heavy Flavor Tracker HFT is a lightweight two layer detector based on Active Pixel Sensors (APS)\cite{APS}, with sensor layers at a radius of 1.5 cm and 5.0 cm, providing a spatial resolution of better than 10 $\mu$m. A fast intermediate device, the Intermediate Silicon Tracker IST, using three layers of standard back-to-back silicon strips and pixel detectors, will act as a pointing device from the TPC to the HFT, and provide high rate capability for high luminosity running. This detector will replace the existing SVT, which does not have sufficient rate capability for future collider luminosities.

\section{Quark Helicities and W Program}

From polarized DIS experiments it is known that the flavor-integrated contribution of the quarks to the proton spin is surprisingly small. A flavor separated study of quark and anti-quark polarizations is thus of fundamental interest to further investigate this question. At RHIC, a flavor separated measurement will be carried out via the maximally parity violating production of $W$ bosons in $u\bar{d} \rightarrow W^+$ and $d\bar{u} \rightarrow W^-$ reactions. This provides a clean access to the quark polarizations, since the $W$ boson couples only to left-handed quarks and right-handed anti-quarks. For $W$ production at large rapidities, the quark is most likely a valence quark from the $p$ traveling in the same direction as the $W$, while the anti-quark comes from the sea of the other $p$. 

At STAR, these $W$s will be detected via their decay $W^+ \rightarrow e^+\nu_e$ and $W^- \rightarrow e^-\bar{\nu}_e$. The energy of the outgoing lepton will be measured in the EEMC at forward rapidity. In order to distinguish between $W^+$ and $W^-$, the identification of the charge sign of these high-momentum electrons is crucial. To achieve this, high resolution tracking in the range $1 < \eta < 2$ is needed. This will be provided by two detector systems. The Forward Silicon Tracker FST will consist of up to 4 silicon disks using conventional back-to-back silicon strip detectors close to the interaction point. The Forward GEM Tracker FGT will provide additional points with a larger lever arm. For this second device, several different geometries, such as two large-area tracking layers in front of the EEMC or two detector barrels within the TPC field cage, are currently under investigation. The FGT will be based on triple-GEM detectors, a technology already successfully applied by the COMPASS experiment\cite{GEMCompass}. For such a large-scale project the commercial availability of GEM foils is crucial. A collaboration with the company TechEtch of Plymouth, MA, USA has been established to provide these foils. First successful tests with detector prototypes have been achieved.

\section{Forward Meson Production}

A significant transverse single spin asymmetry has been observed at large $x_F$ in transversely polarized p+p collisions\cite{Pi0Trans}. So far the data can not discriminate between different models, such as the Sivers and the Collins effect\cite{SpinRev}. While the Sivers effect leads to an asymmetry in forward jet or $\gamma$ production, the Collins mechanism leads to an asymmetry in the forward jet fragmentation. To distinguish between these two scenarios, the jet axis in addition to individual mesons has to be measured. This requires a large acceptance calorimeter in the forward direction. In d+Au collisions, forward meson production is sensitive to the low-$x$ ($0.001<x<0.02$) part of the gluon distribution in the nucleon, probing possible saturation effects of gluon densities.

These measurements will be addressed by an upgraded FPD, the Forward Meson Spectrometer FMS. It is a Pb-glass calorimeter, covering $2.5 < \eta < 4.0$, giving STAR almost hermetic electromagnetic coverage in the range $-1 < \eta < 4$. A first phase of this upgrade, the FPD++, was installed for the 2006 RHIC run.

\section{Summary}

STAR is pursuing a challenging spin physics program that will extend well into the next decade. Future measurements of the gluon polarization via heavy flavor production, the flavor separation of quark and anti-quark polarizations via $W$ production and comprehensive studies of transverse spin phenomena require upgrades of the tracking system and of the electromagnetic calorimetry. The planned tracking upgrade includes high resolution active pixel silicon sensors, standard single-sided silicon strip and pixel detectors and larger area triple-GEM trackers. The upgrade of the forward Pb-glass calorimeter is already well under way. These upgrades will enable STAR to make full use of the polarized proton beams available at RHIC.

\end{document}